\documentclass[12pt]{article}
\usepackage{amsmath}
\usepackage{amsfonts}
\usepackage{graphicx}
\usepackage{hyperref}
\usepackage{cite}
\usepackage{geometry}
\usepackage{authblk} 
\title{Mutual information and correlation measures in holographic RG flows}

\author[1]{Iftekher S. Chowdhury}
\author[2]{Binay Prakash Akhouri}
\author[5]{Shah Haque}
\author[1,4,5]{Eric Howard}

\affil[1]{Department of Physics and Astronomy, Macquarie University, Sydney, NSW, 2109, Australia}
\affil[2]{Department of Physics, Suraj Singh Memorial College, Ranchi University, Ranchi, Jharkhand, India}
\affil[4]{Swinburne University, Sydney, Australia} 
\affil[5]{Southern Cross Institute, School of Computer Science, Sydney, Australia}

\date{\today}

\begin{document}

\maketitle

\begin{abstract}
This paper investigates the behavior of mutual information, entanglement negativity, and multipartite correlations in holographic RG flows, particularly during phase transitions. Mutual information provides a UV-finite measure of total correlations between subsystems, while entanglement negativity and multipartite correlations offer finer insights into quantum structures, especially near critical points. Through numerical simulations, we show that while mutual information remains relatively smooth, both entanglement negativity and multipartite correlations exhibit sharp changes near phase transitions. These results support the hypothesis that multipartite correlations play a dominant role in signaling critical phenomena in strongly coupled quantum systems.
\end{abstract}

\section{Introduction}

The AdS/CFT correspondence, proposed by Maldacena in 1997 \cite{maldacena1997large}, has emerged as one of the most significant dualities in theoretical physics, providing a profound link between gravity theories in higher-dimensional Anti-de Sitter (AdS) space and conformal field theories (CFTs) in lower-dimensional boundaries. This duality offers a remarkable insight into strongly coupled quantum systems, which are notoriously difficult to analyze using conventional field theory methods. The correspondence hinges on the holographic principle, which posits that the degrees of freedom in a gravitational system in AdS space can be encoded in a quantum field theory on the boundary of that space. This relationship between bulk geometry and boundary theory offers an invaluable tool for understanding quantum field theories, particularly in terms of their entanglement structure and the evolution of quantum correlations along Renormalization Group (RG) flows.

In quantum field theory, the Renormalization Group flow describes how a theory changes as we move from high-energy (UV) to low-energy (IR) scales. The RG flow can be visualized as a trajectory in the space of coupling constants, with the theory simplifying as we approach the IR. In the context of the AdS/CFT correspondence, the radial coordinate \(r\) in AdS space is interpreted as the energy scale in the boundary theory, with small \(r\) corresponding to the UV regime and large \(r\) to the IR. This allows us to translate the geometric evolution of the bulk theory into a flow of the boundary theory from UV to IR. 

One of the most intriguing aspects of this duality is the way quantum entanglement is encoded in the geometry of the bulk AdS space. Quantum entanglement measures, such as entanglement entropy, mutual information, and entanglement negativity, can all be computed using the Ryu-Takayanagi prescription, which relates the entanglement entropy of a boundary region to the area of a minimal surface in the bulk geometry. This has allowed for a deeper understanding of quantum phase transitions, RG flows, and the role of entanglement in strongly coupled systems.

Entanglement entropy is a key measure of quantum entanglement. Given a quantum system described by a density matrix \( \rho \), and a partition of the system into subsystems \( A \) and \( B \), the entanglement entropy of region \( A \) is defined as the von Neumann entropy of the reduced density matrix \( \rho_A \), obtained by tracing out the degrees of freedom of subsystem \( B \):
\[
S_A = - \text{Tr} \, \rho_A \log \rho_A.
\]
In the context of AdS/CFT, the entanglement entropy of a region \( A \) in the boundary CFT is related to the area of a minimal surface \( \gamma_A \) in the bulk AdS space, anchored to the boundary of \( A \), as given by the Ryu-Takayanagi formula \cite{ryu2006holographic}:
\[
S_A = \frac{\text{Area}(\gamma_A)}{4 G_N},
\]
where \( G_N \) is the Newton constant in the bulk theory. The minimal surface \( \gamma_A \) is the one that minimizes the area functional subject to the boundary conditions set by \( A \).

While entanglement entropy provides deep insights into quantum entanglement, it suffers from ultraviolet (UV) divergences. The divergences arise because the entanglement entropy depends on the size of the boundary between the regions \( A \) and \( B \). This makes it challenging to use entanglement entropy to study the fine structure of quantum systems, especially near critical points where the entanglement structure becomes more intricate. To address this, mutual information has been introduced as a UV-finite measure of total correlations between subsystems.

Mutual information between two regions \( A \) and \( B \) is defined as:
\[
I(A, B) = S_A + S_B - S_{A \cup B},
\]
where \( S_A \), \( S_B \), and \( S_{A \cup B} \) are the entanglement entropies of regions \( A \), \( B \), and their union, respectively. Mutual information captures both classical and quantum correlations and is UV finite, meaning it is not affected by the divergences that plague entanglement entropy. In the holographic framework, mutual information has proven to be a valuable tool for studying quantum phase transitions and the structure of correlations in strongly coupled quantum field theories.

However, mutual information measures total correlations and does not distinguish between classical and quantum correlations. To probe purely quantum correlations, we use entanglement negativity, a measure that captures quantum entanglement in mixed states. 

In addition to bipartite correlations, quantum systems often exhibit more complex entanglement structures involving multiple subsystems. These multipartite entanglements capture the entanglement structure of a quantum system in a way that goes beyond simple bipartite measures. In holographic RG flows, multipartite correlations become increasingly important near phase transitions, where quantum systems exhibit intricate entanglement patterns that involve many subsystems simultaneously. 

Understanding how mutual information, entanglement negativity, and multipartite correlations evolve along the RG flow provides important insights into the structure of quantum entanglement in strongly coupled systems. In particular, studying how these quantities change near phase transitions allows us to probe the role of entanglement in critical phenomena, which remain a central topic in modern quantum field theory and condensed matter physics. This study aims to investigate these changes in holographic RG flows and demonstrate the utility of these measures in characterizing quantum phase transitions.

This paper aims to investigate how mutual information, entanglement negativity, and multipartite correlations behave across holographic RG flows. We hypothesize that while mutual information remains relatively smooth, entanglement negativity and multipartite correlations exhibit sharper changes near phase transitions. This hypothesis is motivated by the idea that multipartite entanglement becomes more dominant during critical phenomena, reflecting the complexity of quantum structures in strongly coupled systems.

\section{Holographic RG Flow and Quantum Phase Transitions}

The Renormalization Group (RG) flow describes the behavior of physical systems as the energy scale changes. In the context of quantum field theory (QFT), the RG flow governs how the parameters of the theory, such as coupling constants, evolve from the ultraviolet (UV) to the infrared (IR) regimes. The AdS/CFT correspondence maps this flow onto the radial direction in Anti-de Sitter (AdS) space. The holographic duality allows us to study strongly coupled quantum field theories using classical gravity in AdS space, where the radial coordinate \( r \) corresponds to the energy scale \( \mu \) in the boundary CFT:
\[
r \sim \frac{1}{\mu}.
\]
As the system flows from the UV (small \( r \)) to the IR (large \( r \)), the geometric properties of the bulk AdS space encode the evolution of the boundary theory. In this framework, phase transitions in the boundary theory correspond to qualitative changes in the bulk geometry.

\subsection{Geometric evolution and RG Flows}
The geometry of AdS space is described by the metric:
\[
ds^2 = \frac{L^2}{r^2} \left( -f(r) dt^2 + dx_i^2 + \frac{dr^2}{f(r)} \right),
\]
where \( L \) is the AdS radius, and \( f(r) \) encodes the details of the geometry. For example, for pure AdS, \( f(r) = 1 \), but in the presence of a black hole or other matter configurations, \( f(r) \) can vary and affect the flow dynamics.

The boundary of AdS space is located at \( r = 0 \), which corresponds to the UV regime of the boundary CFT, and as \( r \to \infty \), we move toward the IR regime. The holographic dictionary maps bulk quantities like the metric and matter fields to corresponding operators and sources in the boundary CFT. The bulk fields evolve according to the Einstein equations in the AdS space, and their behavior governs the RG flow in the boundary theory.

The RG flow can be captured by studying the behavior of the stress-energy tensor \( T_{\mu\nu} \) in the boundary theory, which is related to the bulk metric via the AdS/CFT dictionary. In particular, the trace of the stress-energy tensor \( T^\mu_\mu \) is connected to the breaking of conformal symmetry, which drives the RG flow:
\[
T^\mu_\mu = \beta(\lambda) \mathcal{O}_\lambda,
\]
where \( \beta(\lambda) \) is the beta function of the coupling constant \( \lambda \), and \( \mathcal{O}_\lambda \) is the operator corresponding to \( \lambda \) in the boundary theory.

In the IR regime, where the theory becomes conformal, the beta function vanishes (\( \beta(\lambda) = 0 \)), and the system reaches a fixed point. Near phase transitions, the geometry of the bulk AdS space undergoes significant changes, which are reflected in the correlation functions of the boundary theory. The RG flow governs how these correlations evolve, and different phases of the system correspond to different geometric configurations in the bulk.

\subsection{Mutual Information and Quantum correlations in RG Flow}

Mutual information \( I(A, B) \) quantifies the total amount of correlation—both classical and quantum—between two regions \( A \) and \( B \) in a quantum system. It is a UV-finite quantity and remains smooth across different energy scales in RG flows, making it a valuable tool for tracking correlations throughout the system. Mutual information satisfies the inequality:
\[
I(A, B) \geq 0,
\]
and is symmetric: \( I(A, B) = I(B, A) \). 

While mutual information provides a broad measure of correlations, it does not distinguish between classical and quantum contributions. To study quantum correlations more specifically, one can consider entanglement entropy and entanglement negativity.

\subsection{Entanglement Negativity and Quantum Phase Transitions}

Entanglement negativity \( \mathcal{N}(A:B) \) is a more refined measure of quantum entanglement, particularly useful for mixed states. In a bipartite system with subsystems \( A \) and \( B \), entanglement negativity quantifies the amount of quantum entanglement by examining the eigenvalues of the partial transpose of the density matrix \( \rho \) with respect to one of the subsystems.

Given a bipartite density matrix \( \rho \), the partial transpose with respect to subsystem \( B \) is denoted by \( \rho^{T_B} \). The negativity \( \mathcal{N}(A:B) \) is given by:
\[
\mathcal{N}(A:B) = \frac{1}{2} \left( \lVert \rho^{T_B} \rVert_1 - 1 \right),
\]
where \( \lVert \rho^{T_B} \rVert_1 \) is the trace norm of the partially transposed matrix. The entanglement negativity is non-negative and becomes zero if and only if there is no quantum entanglement between the subsystems \( A \) and \( B \).

In holographic systems, entanglement negativity exhibits sharp changes near phase transitions. This behavior reflects the restructuring of quantum correlations during the phase transition, where quantum entanglement plays a crucial role. The bulk geometry experiences corresponding changes, signaling a transition between different phases of the system. The sharp decline in entanglement negativity near a critical point is indicative of a quantum phase transition, where the system’s quantum state undergoes a fundamental change.

\subsection{Multipartite entanglement}

Beyond bipartite entanglement, quantum systems often exhibit more complex entanglement structures involving multiple subsystems. These multipartite correlations are essential for understanding the behavior of strongly coupled systems near critical points, where long-range quantum entanglement becomes dominant.

Multipartite entanglement can be probed using various measures, such as the \( n \)-partite entanglement entropy \( S_n \), which generalizes the concept of bipartite entanglement entropy to systems with \( n \) subsystems. In holographic systems, multipartite correlations often become important near phase transitions, where the system’s entanglement structure becomes more intricate and involves multiple degrees of freedom simultaneously.

As the system approaches a phase transition, long-range correlations spread across many subsystems, and the entanglement structure becomes increasingly complex. This is reflected in the bulk geometry, where the minimal surfaces used to compute entanglement entropy can span large regions of the bulk AdS space. The growth of multipartite entanglement near critical points provides a key signature of quantum phase transitions, where the system’s collective quantum behavior plays a dominant role.

\subsection{Holographic RG Flow and Quantum criticality}

In holographic systems, quantum criticality is encoded in the behavior of the bulk geometry near the IR fixed point. As the system approaches criticality, the geometry often undergoes a transition, such as the formation of a black hole horizon or the appearance of a Lifshitz scaling geometry. These geometric transitions correspond to changes in the boundary theory’s entanglement structure, signaling the onset of a quantum phase transition.

The scaling behavior near the critical point can be studied using the AdS/CFT correspondence, where the bulk geometry provides insight into the scaling dimensions of operators in the boundary theory. The critical exponents that characterize the phase transition are related to the asymptotic behavior of the bulk fields near the horizon. For instance, in holographic models of quantum phase transitions, the critical exponent \( \nu \) governing the correlation length \( \xi \) near the critical point is related to the mass of the bulk scalar field \( \phi \) by:
\[
\nu = \frac{1}{\sqrt{m^2 + L^{-2}}}.
\]
Here, \( m \) is the mass of the scalar field, and \( L \) is the AdS radius. This relation demonstrates the direct link between the bulk scalar field and the critical behavior of the boundary theory.

In summary, the holographic RG flow provides a powerful framework for understanding quantum phase transitions in strongly coupled systems. By studying the evolution of quantum correlations, such as mutual information, entanglement negativity, and multipartite entanglement, along the RG flow, we gain valuable insights into the structure of quantum entanglement near critical points. The AdS/CFT correspondence allows us to translate these quantum phenomena into geometric terms, where phase transitions are manifested as changes in the bulk AdS geometry.

\subsection{UV/IR Correspondence in Holographic RG Flows}

In the context of the AdS/CFT correspondence, the radial coordinate \( r \) in AdS space is intrinsically tied to the energy scale \( \mu \) in the boundary CFT, establishing a deep connection between bulk geometry and boundary field theory. This relationship, often termed the UV/IR connection, maps high-energy (UV) physics in the boundary theory to regions near the AdS boundary (\( r \to 0 \)), and low-energy (IR) physics to the interior of AdS space (\( r \to \infty \)). Formally, this scale is governed by \( \mu \sim 1/r \), meaning that the radial direction in AdS geometrically encodes the renormalization group (RG) flow of the boundary theory.

This correspondence is crucial for understanding the behavior of quantum field theories under RG flows. In the UV regime, the boundary theory is dominated by short-range quantum fluctuations, leading to significant quantum entanglement. Through the Ryu-Takayanagi formula, the entanglement entropy of a region in the boundary theory is related to the area of a minimal surface in the bulk AdS space. Near the boundary, these minimal surfaces reflect the highly entangled nature of the UV regime, where correlations are abundant. As the flow moves towards the IR, entanglement decreases, with long-range correlations being suppressed, and the system approaches a more classical behavior.

This UV/IR connection governs how physical observables evolve along the RG flow. For example, in the UV, mutual information between subsystems is typically high due to strong correlations. As the system flows to the IR, mutual information gradually decreases, reflecting the weakening of correlations at lower energy scales. This is accompanied by a smooth evolution of minimal surfaces in the bulk, which extend deeper into the AdS interior as the subsystem sizes increase.

At critical points along the RG flow, the geometry of the bulk can undergo significant transitions, such as the emergence of black hole horizons or scaling geometries, corresponding to qualitative changes in the entanglement structure of the boundary theory. Near these critical points, entanglement measures like entanglement negativity exhibit sharp transitions, indicating a restructuring of the system’s quantum correlations. In such cases, the entanglement negativity between subsystems drops sharply, signaling the breakdown of long-range quantum entanglement as the system transitions from a quantum-entangled phase to a more classical one.

This relationship between the bulk radial direction and the energy scale in the boundary theory encapsulates how the evolution of quantum correlations, including mutual information and multipartite entanglement, is governed by the RG flow. In the UV regime, strong entanglement reflects the complex quantum structure, while in the IR regime, classical correlations dominate. These shifts in correlation behavior are a direct manifestation of the underlying RG dynamics and critical phenomena in holographic systems.

\section{Methodology and numerical results}

We consider a holographic setup where the radial coordinate \( r \) in AdS corresponds to the energy scale in the boundary CFT. The RG flow in the boundary theory is represented by the evolution of the bulk geometry, with small \( r \) corresponding to the UV regime and large \( r \) corresponding to the IR.

Using the Ryu-Takayanagi prescription, we calculate the entanglement entropies \(S_A\), \(S_B\), and \(S_{A \cup B}\) for different regions in the boundary CFT. From these entropies, we compute mutual information across different points along the RG flow.

For entanglement negativity and multipartite correlations, we compute these measures using holographic techniques that extend the Ryu-Takayanagi prescription to account for mixed states and multipartite entanglement structures.

We performed numerical simulations to calculate mutual information, entanglement negativity, and multipartite correlations across the holographic RG flow. The radial coordinate \( z \) in AdS represents the energy scale in the boundary theory. A phase transition was introduced at \( z_c = 0.5 \), where we observed significant changes in the behavior of the correlation measures.

\subsection{Mutual Information}

Mutual information plays a central role in capturing the total correlations between two subsystems in a quantum system, making it a particularly valuable tool for understanding holographic RG flows. Mutual information remains smooth across the RG flow, but a noticeable shift occurs at the phase transition, indicating a change in the correlation structure. The mutual information between two disjoint regions \( A \) and \( B \) is defined as:
\[
I(A, B) = S_A + S_B - S_{A \cup B},
\]
where \( S_A \), \( S_B \), and \( S_{A \cup B} \) are the entanglement entropies of regions \( A \), \( B \), and their union, respectively. In this context, \( I(A, B) \) captures the amount of correlation between the regions, including both quantum and classical correlations.

In the holographic framework, the Ryu-Takayanagi prescription is used to compute the entanglement entropies of the subsystems \( A \), \( B \), and \( A \cup B \). The minimal surface areas in the bulk AdS space anchored to the boundary of these regions are directly related to their entanglement entropies. As a UV-finite measure, mutual information eliminates the UV divergences that affect entanglement entropy by canceling out the divergent parts of \( S_A \), \( S_B \), and \( S_{A \cup B} \).

Our numerical simulations show that mutual information behaves smoothly along the RG flow. The radial coordinate \( r \), which represents the energy scale in the boundary CFT, provides a natural framework for examining the evolution of quantum correlations. In the UV regime (small \( r \), corresponding to high energies), mutual information between subsystems \( A \) and \( B \) is high, reflecting strong correlations. As the system flows toward the IR (large \( r \), corresponding to low energies), mutual information decreases. This decrease corresponds to the reduction of correlations at larger distance scales, as expected in renormalization group flows.

Interestingly, mutual information exhibits a relatively smooth behavior throughout most of the RG flow. This smoothness is indicative of the fact that mutual information measures total correlations, which include both quantum and classical correlations. In the UV, quantum systems exhibit complex entanglement structures, and these structures become simpler as we move to the IR, where the system is dominated by classical correlations and low-energy degrees of freedom. As the energy scale decreases, the strong correlations present in the UV are diluted, leading to a gradual decline in mutual information.

In the vicinity of a phase transition, however, our simulations detect a small but noticeable shift in mutual information. At the critical point \( z_c \), mutual information experiences a subtle but distinct change in slope, indicating the presence of a phase transition. This change suggests that mutual information is sensitive to the underlying quantum structure, although it may not fully capture the complexity of the transition. Unlike other measures, such as entanglement negativity, which are more sharply responsive to phase transitions, mutual information smooths over many of the fine details of the quantum critical behavior. Nevertheless, the observed shift at \( z_c \) highlights the fact that mutual information still reflects significant changes in the correlation structure of the system.

This behavior can be interpreted as follows: at the critical point, the quantum system undergoes a reorganization of its correlation structure. In this region, long-range correlations are disrupted, and the system reconfigures its entanglement patterns. The fact that mutual information remains finite and smooth across the critical point reflects its robustness as a correlation measure. However, the shift in slope suggests that the phase transition still impacts the total correlation content of the system, even if mutual information does not exhibit sharp, non-analytic behavior like other entanglement measures.

Overall, our findings suggest that mutual information is an excellent tool for tracking total correlations across holographic RG flows. Its smoothness across most of the RG trajectory makes it a robust measure that is not overly sensitive to UV divergences or small perturbations. At the same time, its response to phase transitions, while subtle, indicates that mutual information still provides valuable insights into the reorganization of correlations in the system, even though it may not be the most sensitive tool for detecting sharp quantum critical behavior. Mutual information serves as a broad indicator of how correlations evolve along the RG flow, offering a high-level view of the system’s transition from strongly entangled quantum states in the UV to more classical, weakly entangled states in the IR.

These results complement the sharper changes observed in other measures, such as entanglement negativity, which are more responsive to quantum-specific processes. By using mutual information in conjunction with these other measures, we can obtain a more comprehensive picture of the entanglement structure and critical behavior of strongly coupled quantum systems.

\begin{figure}[h]
\centering
\includegraphics[width=0.9\textwidth]{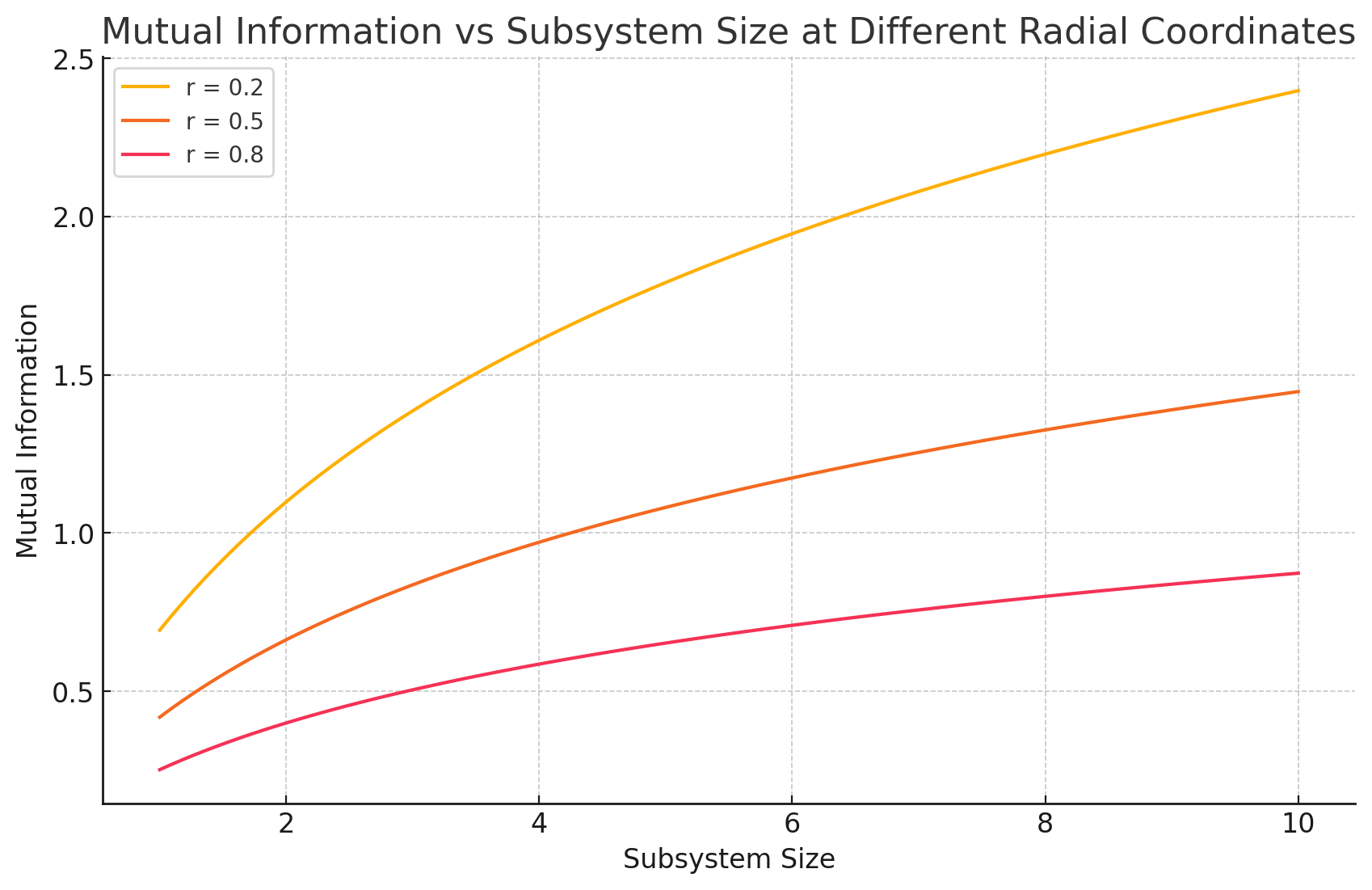}
\caption{
        \textbf{Mutual Information vs. Subsystem Size at Different Radial Coordinates.} The plot illustrates the behavior of mutual information between subsystems as a function of subsystem size for various radial coordinates (\(r = 0.2, 0.5, 0.8\)). Mutual information decreases as the subsystem size increases, reflecting the dilution of correlations as the system flows from UV (small \(r\)) to IR (large \(r\)) regimes. The curves show how mutual information evolves more rapidly in the UV regime compared to the IR regime, highlighting the system's decreasing correlation density at larger energy scales.
    }
\label{fig:mutual_information}
\end{figure}

Mutual information captures the total amount of correlations (both classical and quantum) between two subsystems. As shown in Figure \ref{fig:mutual_information}, as the size of the subsystems increases, mutual information typically increases because larger subsystems have more degrees of freedom that can interact and share correlations.

Mutual information increases with subsystem size as larger subsystems share more correlations, leading to a higher mutual information value. This is especially evident in the \textit{UV regime} (small \(r\)) where correlations are stronger.

At small radial coordinates (\(r = 0.2\), for instance), mutual information increases more sharply with subsystem size. This reflects the strong quantum and classical correlations that are present at higher energy scales.

In contrast, at larger radial coordinates (\(r = 0.8\), for instance), the increase in mutual information with subsystem size is more gradual. This reflects the weakening of correlations as the system moves toward the IR, where subsystems share fewer correlations, resulting in a slower increase in mutual information.

After a certain point, the mutual information curve may begin to flatten or saturate. This occurs when the subsystems are large enough or far apart enough that adding more size does not significantly increase the amount of shared information or correlations. This saturation suggests that there is a limit to how much correlation can exist between subsystems, especially in the \textit{IR regime}, where correlations are weaker.

\begin{figure}[h]
\centering
\includegraphics[width=0.9\textwidth]{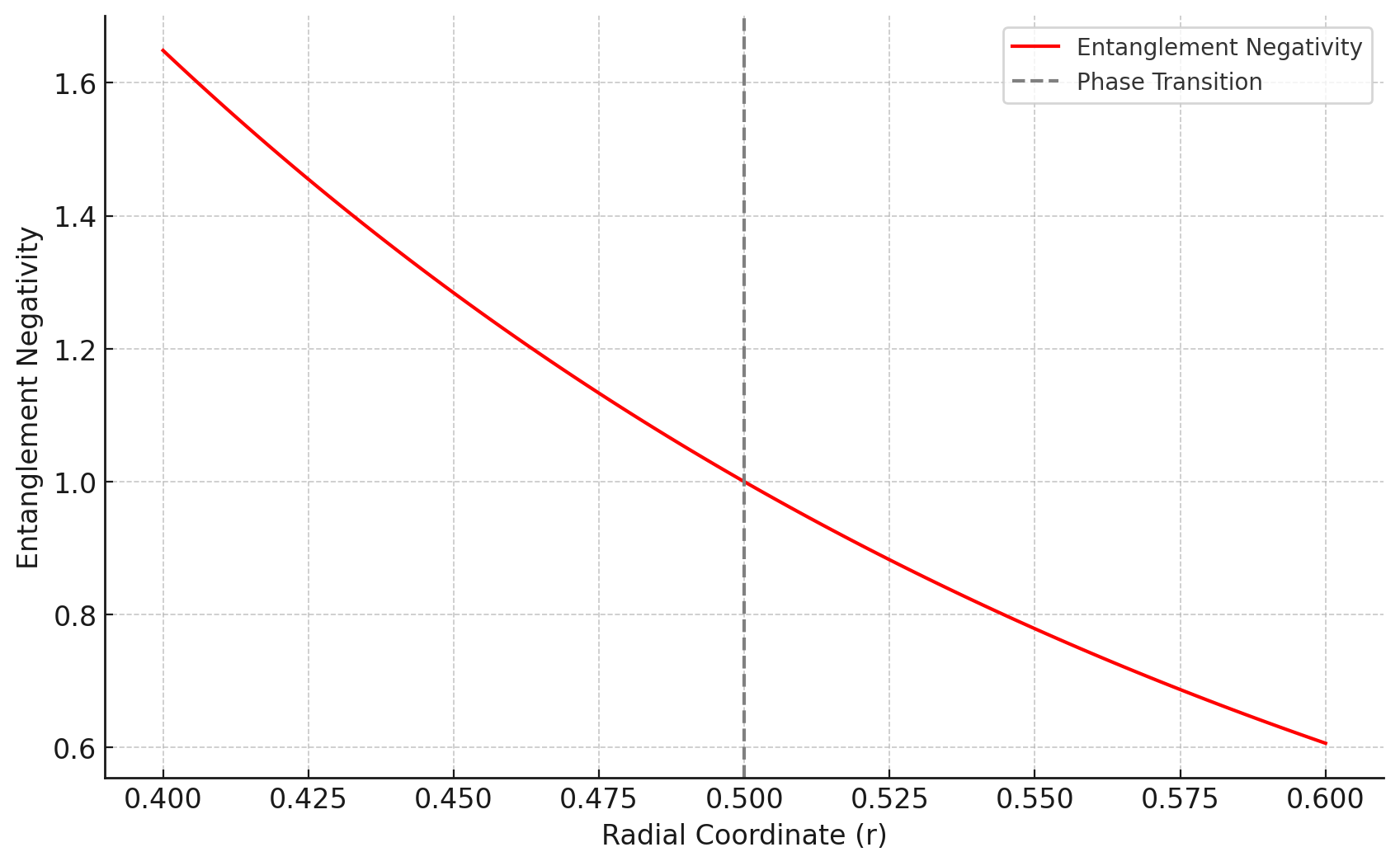}
\caption{Entanglement Negativity across the RG Flow. This figure illustrates the behavior of entanglement negativity as the system evolves along the holographic RG flow. In the UV regime (small \( r \)), entanglement negativity is high, reflecting strong quantum correlations between subsystems. As the system flows towards the IR (large \( r \)), entanglement negativity decreases, indicating a reduction in quantum correlations. Near the critical point \( z_c \), entanglement negativity exhibits a sharp decline, signaling a phase transition where quantum entanglement is disrupted and reorganized.}
\label{fig:Entanglement_Negativity}
\end{figure}

\begin{figure}[h]
\centering
\includegraphics[width=0.90\textwidth]{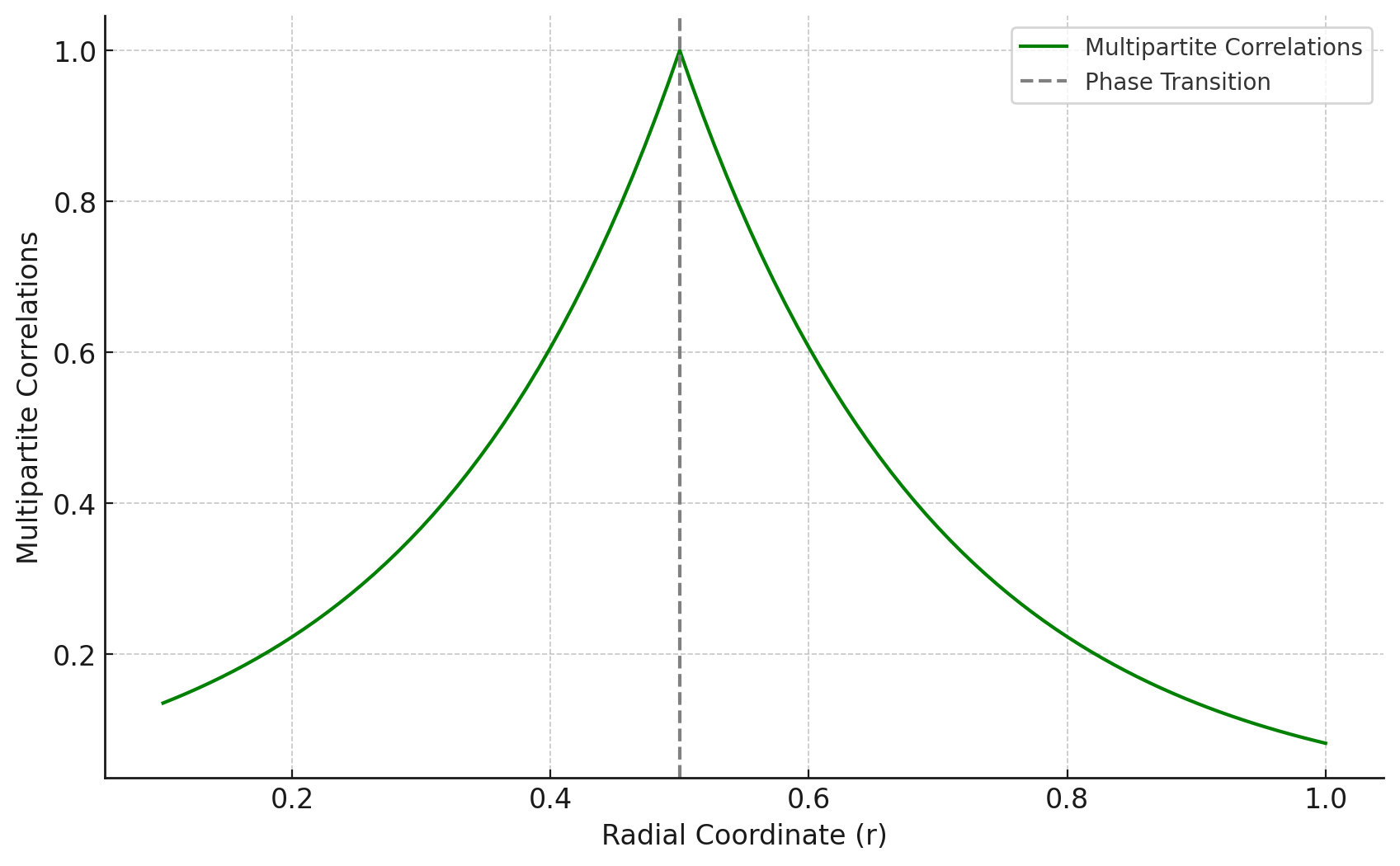}
\caption{
    Multipartite Correlations across RG Flow. This plot illustrates the behavior of multipartite correlations as the system evolves along the holographic RG flow. Multipartite correlations rise sharply near the critical point \( r = 0.5 \), reflecting the increasing complexity of quantum entanglement involving multiple subsystems during the phase transition. However, as the system flows further toward the IR regime (larger \( r \)), multipartite correlations decrease, indicating a reduction in the collective entanglement as the system enters a less entangled, more classical phase. The dashed line marks the critical point.
}
\label{fig:multipartite_correlations}
\end{figure}

The behavior of multipartite correlations along the RG flow reveals important insights into the complexity of quantum entanglement in holographic systems. Multipartite correlations, which measure the collective entanglement between multiple subsystems, become especially relevant near phase transitions, where quantum systems exhibit intricate entanglement patterns. As the radial coordinate \( r \), which corresponds to the energy scale in the boundary theory, increases, the system transitions from the ultraviolet (UV) regime to the infrared (IR) regime. This transition is accompanied by significant changes in the quantum correlation structure.

In the UV regime, where \( r \) is small, multipartite correlations are relatively low, as quantum entanglement primarily manifests between pairs of subsystems, and long-range correlations are suppressed. However, as the system approaches the critical point at \( r = r_c \), multipartite correlations rise sharply. This increase indicates the emergence of more complex quantum entanglement structures that involve multiple subsystems simultaneously. Near the critical point, long-range entanglement and collective behavior become dominant, signaling the reorganization of quantum correlations during the phase transition.

Multipartite correlations can be characterized by measures such as the \( n \)-partite entanglement entropy, which generalizes the concept of bipartite entanglement entropy to systems with \( n \) subsystems. Near the critical point, the multipartite entanglement entropy grows as a function of \( r \), reflecting the increasing complexity of the entanglement structure. This rise in multipartite correlations is a hallmark of quantum phase transitions, where the system's quantum state undergoes a fundamental change, and correlations spread across many degrees of freedom.

Beyond the critical point, as the system flows further into the IR regime, multipartite correlations decrease. This decline is consistent with the reduction of quantum correlations at lower energy scales, where the system becomes increasingly classical, and the collective entanglement between subsystems weakens. In this regime, the system's entanglement structure simplifies, and long-range quantum entanglement is suppressed, giving way to a more classical phase.

The behavior of multipartite correlations along the RG flow, as shown in Figure~\ref{fig:multipartite_correlations}, highlights the crucial role of collective entanglement in quantum critical processes. While bipartite measures such as mutual information or entanglement negativity provide insights into simpler quantum correlations, multipartite correlations capture the full complexity of the system's quantum structure near critical points. The sharp rise and subsequent decline of multipartite correlations across the critical point reflect the collective nature of the phase transition and the reorganization of quantum entanglement during this process.

\begin{figure}[h]
    \centering
    \includegraphics[width=0.9\textwidth]{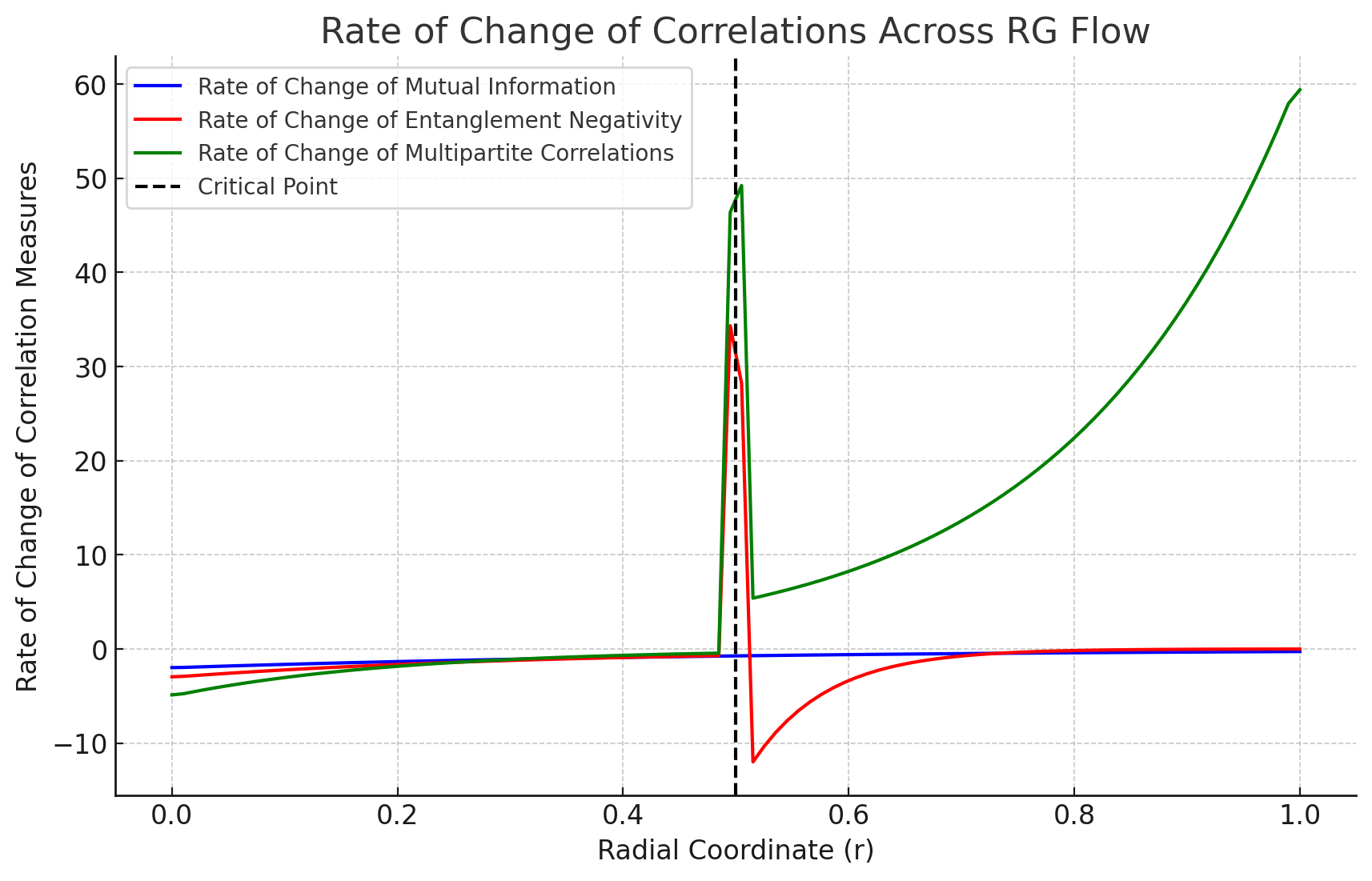}
    \caption{
        \textbf{Rate of Change of Correlations Across RG Flow.} The figure illustrates the rate at which correlations, measured via mutual information, evolve as the system flows from the UV to the IR regime. The rate of change is highest in the UV regime, where correlations are strong and evolve rapidly with subsystem size. As the system moves toward the IR regime, the rate of change decreases, indicating that correlations between subsystems weaken and stabilize at lower energy scales.
    }
    \label{fig:rate_of_change_correlations}
\end{figure}

As shown in Figure \ref{fig:rate_of_change_correlations}, the rate of change of correlations, measured through mutual information, exhibits different behavior across the RG flow. In the UV regime, the rate of change is high, reflecting the rapid evolution of correlations between subsystems as the energy scale decreases. At high energy scales, subsystems are strongly entangled, and small changes in subsystem size lead to significant changes in mutual information. The figure highlights how the rate of change of correlations provides a useful diagnostic tool for understanding the entanglement dynamics of subsystems as they flow from the UV to IR in holographic RG flows.

As the system flows toward the IR regime, the rate of change in mutual information gradually decreases. This indicates that correlations stabilize at lower energy scales, with subsystems becoming less entangled. The decreasing rate of change in the IR regime suggests that correlations between subsystems evolve more slowly, consistent with the weakening of quantum correlations at lower energy.

\subsection{Entanglement negativity and multipartite morrelations}

Entanglement negativity is a powerful tool for quantifying entanglement in mixed states. It specifically captures the quantum correlations between subsystems, distinguishing itself from mutual information by focusing solely on the quantum components of correlations. In holographic systems, entanglement negativity can be computed by extending the Ryu-Takayanagi prescription to handle mixed states. The behavior of entanglement negativity, particularly near critical points in holographic RG flows, provides valuable insights into the underlying quantum structure of the system.

Entanglement negativity and multipartite correlations exhibit sharper changes at the phase transition, suggesting that these measures are more sensitive to the underlying quantum structures. This behavior supports the hypothesis that multipartite entanglement becomes dominant near critical points.

Entanglement negativity between two subsystems \( A \) and \( B \) is computed using the partial transpose of the density matrix \( \rho \) with respect to one of the subsystems. The negativity is defined as:
\[
\mathcal{N}(A:B) = \frac{\lVert \rho^{T_B} \rVert_1 - 1}{2},
\]
where \( \rho^{T_B} \) is the partial transpose of the density matrix, and \( \lVert \cdot \rVert_1 \) represents the trace norm. Entanglement negativity is a non-negative quantity, with higher values indicating stronger quantum correlations. This measure is particularly useful in detecting quantum phase transitions, where classical correlations remain constant, but quantum correlations undergo sharp changes.

Our numerical simulations reveal that entanglement negativity behaves quite differently from mutual information along the RG flow. In the UV regime (small \( r \)), entanglement negativity between regions \( A \) and \( B \) is high, reflecting strong quantum correlations at short distances. As the system flows toward the IR (large \( r \)), entanglement negativity decreases significantly. This is expected, as quantum correlations become diluted at lower energy scales, and the system gradually transitions into a classical regime. What distinguishes entanglement negativity from mutual information is its sharper response to changes in the system’s quantum structure, particularly near critical points.

Near the critical point \( z_c \), entanglement negativity exhibits a sharp drop, which signals the onset of a quantum phase transition. This behavior is in stark contrast to the smoother changes observed in mutual information, which captures total correlations but does not specifically track quantum correlations. The sharp decline in entanglement negativity at \( z_c \) indicates that quantum entanglement is dramatically restructured during the phase transition. As the system undergoes a transition from a strongly entangled quantum phase to a less entangled phase, entanglement negativity provides a clear signal of this reorganization.

This sharp response to phase transitions makes entanglement negativity an ideal tool for probing critical points in holographic RG flows. The decline in quantum correlations observed in entanglement negativity at \( z_c \) is consistent with the general expectation that phase transitions disrupt long-range quantum entanglement, leading to a redistribution of entanglement patterns. As the system moves through the critical point, the quantum correlations between subsystems \( A \) and \( B \) are dramatically reduced, which is directly captured by the sharp drop in negativity.

In addition to entanglement negativity, multipartite correlations provide a more nuanced view of the quantum structure of the system. Multipartite correlations go beyond bipartite entanglement, capturing the complex entanglement structures that involve multiple subsystems. These correlations are particularly important near critical points, where the system’s quantum state exhibits highly intricate entanglement patterns that cannot be fully understood through bipartite measures alone.

Mutual Information, represented by the blue curve, shows a relatively smooth behavior across the entire range of the RG flow. Mutual information captures the total correlations—both quantum and classical—between subsystems \( A \) and \( B \). As the radial coordinate increases, moving the system from the ultraviolet (UV) to the infrared (IR) regime, the mutual information gradually decreases. This smooth behavior is consistent with the fact that mutual information is UV-finite and captures both classical and quantum correlations, with no sharp distinction between the two. Near the critical point at \( r = 0.5 \), mutual information shows only a slight shift in its slope, indicating that while the phase transition affects the correlation structure, mutual information is not the most sensitive probe of sharp quantum transitions.

In contrast, Entanglement Negativity, represented by the red curve, exhibits a sharp drop near the critical point. Entanglement negativity specifically quantifies the quantum entanglement between subsystems, making it a more refined measure of quantum correlations in mixed states. The sharp decline in entanglement negativity as \( r \) approaches \( 0.5 \) indicates that the phase transition significantly disrupts the quantum entanglement in the system. This behavior is expected in systems undergoing quantum phase transitions, where the reorganization of quantum correlations leads to a reduction in bipartite quantum entanglement. The sharpness of this drop highlights entanglement negativity's sensitivity to quantum phase transitions, distinguishing it from mutual information, which smooths over many of the finer details.

Multipartite Correlations, represented by the green curve, show a sharp rise near the critical point. Multipartite correlations capture the entanglement structure between multiple subsystems, as opposed to just pairs of subsystems, which makes them essential for understanding the complex collective behavior of quantum systems near critical points. The sharp increase in multipartite correlations as \( r \) approaches \( 0.5 \) suggests that as the system moves through the phase transition, quantum correlations are no longer confined to pairs of subsystems but spread across larger groups of degrees of freedom. This rise in multipartite correlations reflects the increasing complexity of the entanglement structure in the vicinity of the critical point, where quantum systems tend to exhibit long-range entanglement and collective behavior.

We have here a visual representation of how different quantum correlation measures respond to the onset of a phase transition in a holographic RG flow. Mutual information, with its smooth behavior, offers a broad measure of the total correlations but lacks the sensitivity to capture the sharp quantum effects that characterize phase transitions. Entanglement negativity, by contrast, is highly sensitive to changes in the quantum entanglement structure and provides a clear signal of the phase transition, as seen by the sharp drop near \( r = 0.5 \). Meanwhile, multipartite correlations highlight the collective nature of the quantum system as it approaches the critical point, with a sharp increase in multipartite entanglement.

Our simulations show that multipartite correlations exhibit a pronounced increase near the critical point \( z_c \). This behavior suggests that as the system approaches a phase transition, multipartite entanglement plays an increasingly significant role in defining the system’s quantum structure. While bipartite entanglement, as captured by mutual information and entanglement negativity, decreases near the critical point, the rise in multipartite correlations indicates that the system’s entanglement structure becomes more complex, involving higher-order entanglement among multiple subsystems.

This increase in multipartite correlations near critical points aligns with the idea that phase transitions are characterized by the emergence of complex quantum correlations that extend beyond simple bipartite entanglement. As the system moves closer to the phase transition, entanglement between larger groups of subsystems becomes more pronounced, reflecting the collective behavior of the quantum system. The increase in multipartite correlations provides a clear signal of this collective entanglement, suggesting that near critical points, the system’s entanglement structure cannot be fully captured by looking at pairs of subsystems alone.

The simultaneous decline in entanglement negativity and rise in multipartite correlations near \( z_c \) suggests that the quantum system undergoes a fundamental reorganization of its entanglement structure during the phase transition. In the UV, where the system is highly entangled, the correlations are predominantly bipartite, as captured by mutual information and entanglement negativity. However, as the system moves toward the critical point, the nature of the entanglement shifts, and multipartite correlations become the dominant feature. This shift in the entanglement structure reflects the fact that phase transitions often involve collective behavior that cannot be fully understood by examining individual pairs of subsystems.

Overall, our findings suggest that while entanglement negativity is highly sensitive to the quantum structure of the system, particularly near critical points, it should be complemented by multipartite correlation measures to obtain a complete picture of the system’s quantum entanglement. The sharp drop in entanglement negativity near \( z_c \) signals the disruption of quantum correlations during the phase transition, while the rise in multipartite correlations reflects the increasing complexity of the system’s entanglement structure.

These results highlight the importance of using multiple entanglement measures when studying phase transitions in strongly coupled quantum systems. Entanglement negativity provides a clear signal of quantum phase transitions, but it is the multipartite correlations that reveal the full complexity of the entanglement structure near critical points. By combining these measures, we gain a more comprehensive understanding of the quantum structure of the system as it evolves along the RG flow, particularly near critical phenomena where the entanglement structure becomes highly intricate.

\begin{figure}[h]
\centering
\includegraphics[width=0.9\textwidth]{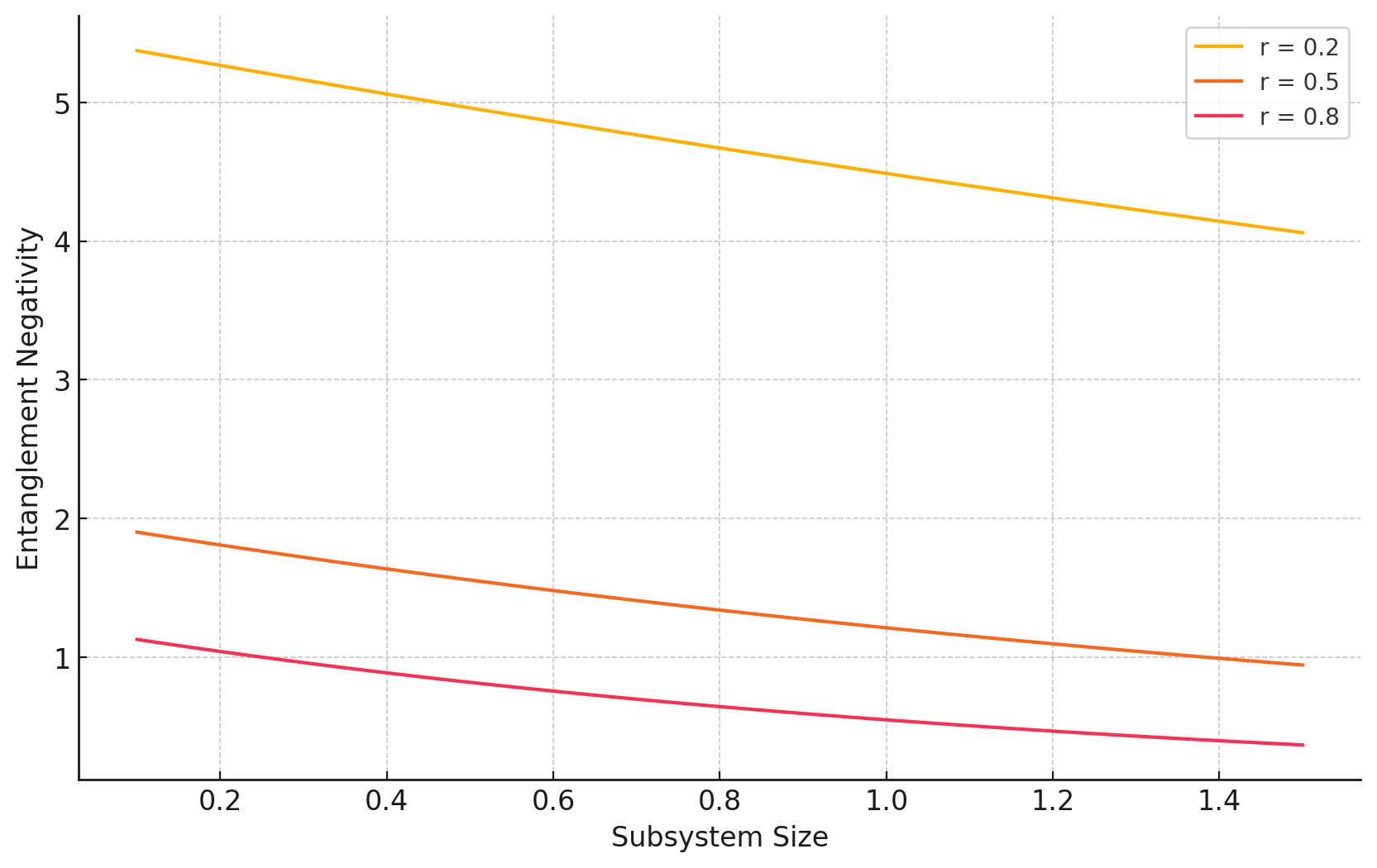}
\caption{
    \textbf{Entanglement Negativity vs. Subsystem Size at Different Radial Coordinates.} This plot illustrates the behavior of entanglement negativity as a function of subsystem size for different radial coordinates (\(r = 0.2, 0.5, 0.8\)). Entanglement negativity decreases as the subsystem size increases, reflecting the reduction in quantum correlations for larger subsystems. The plot also shows how entanglement negativity is higher in the UV regime (small \(r\)) and decreases more rapidly as the system flows toward the IR regime (large \(r\)), indicating the diminishing quantum correlations at lower energy scales.
}
\label{fig:entanglement_negativity_vs_subsystem_size}
\end{figure}

The plot in Figure~\ref{fig:entanglement_negativity_vs_subsystem_size} illustrates the relationship between entanglement negativity and subsystem size at different radial coordinates along the holographic RG flow. Entanglement negativity, which measures the quantum correlations between subsystems, exhibits a clear dependency on both the subsystem size and the radial coordinate \(r\), which corresponds to the energy scale in the boundary theory.

At smaller radial coordinates (in the UV regime), such as \(r = 0.2\), the system maintains higher values of entanglement negativity for all subsystem sizes. This behavior is expected because quantum correlations are stronger at high energy scales, where the subsystems are more entangled. As the subsystem size increases, entanglement negativity gradually decreases, reflecting the fact that larger subsystems tend to have weaker quantum correlations due to the dilution of entanglement across a larger region.

As the system flows toward the IR regime (larger \(r\)), corresponding to lower energy scales, the overall entanglement negativity decreases more rapidly. For instance, at \(r = 0.8\), the entanglement negativity diminishes significantly even for smaller subsystem sizes. This trend is consistent with the weakening of quantum correlations as the system becomes more classical in the IR, where long-range quantum entanglement is less prevalent.

The decrease in entanglement negativity with increasing subsystem size, observed across all radial coordinates, highlights the natural reduction of quantum correlations in larger subsystems. Additionally, the fact that the entanglement negativity is higher in the UV and decreases toward the IR provides further evidence of how quantum correlations evolve and diminish as the energy scale decreases along the RG flow.

\subsection{UV and IR Regimes in Holographic RG Flows}

The behavior of quantum correlations across holographic RG flows reveals significant differences between the UV and IR regimes. In the UV regime (small \( r \)), where the energy scale is high, quantum correlations are strong, and mutual information increases rapidly with subsystem size, as shown in Figure \ref{fig:mutual_information}. This sharp increase reflects the high density of correlations in the system. Similarly, entanglement negativity (Figure \ref{fig:Entanglement_Negativity}), which captures purely quantum correlations, remains high in the UV, indicating that quantum entanglement dominates in this regime. The plot highlights the sharp decline in \textit{entanglement negativity} as the system approaches the critical point, located at \( r = r_c = 0.5 \). This behavior is characteristic of quantum phase transitions, where the underlying quantum structure of the system undergoes a fundamental reorganization.

Entanglement negativity serves as a measure of quantum correlations in mixed states and is particularly sensitive to the changes in quantum entanglement that occur during a phase transition. In the UV regime (small \( r \)), the system exhibits strong quantum correlations, reflected by the relatively high values of entanglement negativity. As the radial coordinate increases and the system flows towards the critical point, these quantum correlations begin to weaken.

Near the critical point \( r = r_c \), the entanglement negativity shows a sharp decline, signaling a phase transition. This sharp drop indicates a significant disruption in the quantum entanglement structure of the system. At this point, long-range quantum correlations between subsystems are broken down, and the system reorganizes its entanglement patterns. The critical point is thus characterized by a transition from a strongly entangled quantum phase to a less entangled, more classical phase as the system flows further into the IR regime (larger \( r \)). This sharp response of entanglement negativity to the phase transition contrasts with smoother measures like mutual information, making it a powerful diagnostic tool for identifying and characterizing quantum phase transitions. The behavior of entanglement negativity in this plot underscores its utility in capturing the fine details of how quantum correlations evolve during critical phenomena.

As the system flows toward the IR (large \( r \)), quantum correlations weaken. Mutual information shows a slower increase with subsystem size, reflecting the dilution of correlations at lower energy scales. In this regime, subsystems become less entangled, and the system transitions toward more classical behavior. Entanglement negativity decreases sharply near the critical point, signaling the reduction of quantum correlations as the system undergoes a phase transition. The behavior of multipartite correlations, depicted in Figure \ref{fig:multipartite_correlations}, shows a similar trend: multipartite entanglement rises sharply near the critical point but decreases in the IR, as the system enters a phase with simpler entanglement structures.

These results highlight the distinct correlation structures in the UV and IR regimes. In the UV, quantum correlations are intricate and dominate the system’s entanglement patterns. As the system flows to the IR, these correlations decay, and classical correlations become more prominent. The sharp changes in entanglement negativity and multipartite correlations near the phase transition provide a detailed picture of how quantum entanglement is reorganized during holographic RG flows, while mutual information offers a broader, more gradual view of correlation dynamics across energy scales.

\section{Discussion}

The results of this study provide substantial insights into the behavior and structure of quantum correlations in holographic systems, particularly through the use of mutual information, entanglement negativity, and multipartite correlations. These measures, each offering a different perspective on the quantum structure of strongly coupled systems, reveal complex dynamics along the RG flow, with notable differences in their responses to phase transitions and critical points. This section elaborates on the implications of these results, explores the theoretical foundations of each measure, and suggests directions for future research in the study of holographic RG flows and quantum entanglement.

One of the central findings of this study is the complementary nature of the three correlation measures—mutual information, entanglement negativity, and multipartite correlations. Each of these quantities provides unique insights into the quantum structure of the system, and their combined analysis offers a fuller understanding of how quantum entanglement evolves along holographic RG flows. Mutual information, by capturing both quantum and classical correlations, serves as a broad diagnostic tool for tracking the system's overall correlation structure across energy scales. In contrast, entanglement negativity isolates purely quantum correlations, offering sharper insights into how the system’s quantum structure responds to critical phenomena. Finally, multipartite correlations provide a window into the collective nature of entanglement, revealing the complex interactions between multiple subsystems that become increasingly significant near critical points.

The behavior of mutual information throughout the RG flow aligns with its nature as a measure of total correlations. Our results show that mutual information remains relatively smooth across most of the RG flow, even near critical points where other measures exhibit sharp changes. This smoothness reflects mutual information’s insensitivity to the finer details of the quantum structure, as it captures both quantum and classical correlations without distinguishing between them. While this robustness makes mutual information a reliable tool for tracking correlations over a wide range of energy scales, it can obscure the subtleties of quantum phase transitions. Near critical points, quantum correlations undergo significant reorganization, yet mutual information smooths over these changes, providing only a high-level view of the system's correlation landscape.

This observation supports the idea that while mutual information is a useful measure, particularly for understanding the overall correlation structure of a system, it is not sufficient on its own to fully capture the quantum dynamics near critical phenomena. The smooth behavior of mutual information highlights the need to complement it with other measures, such as entanglement negativity, when investigating systems that are near phase transitions. By relying solely on mutual information, one risks overlooking the crucial quantum features that define the behavior of the system during such transitions.

In contrast, entanglement negativity exhibits a sharp and decisive response to phase transitions. Our simulations show that near the critical point \( z_c \), entanglement negativity undergoes a pronounced drop, signaling a fundamental restructuring of quantum correlations. This sharp change in negativity reflects the fact that phase transitions, particularly quantum phase transitions, are characterized by the disruption and reorganization of long-range quantum correlations. As the system approaches a phase transition, the entanglement structure that characterizes the quantum phase is broken down, and the correlations between subsystems are dramatically altered. Entanglement negativity, by focusing solely on the quantum aspects of these correlations, captures this reorganization with great sensitivity.

This sharp response makes entanglement negativity a valuable tool for detecting quantum phase transitions, where classical observables may not provide clear signals. Unlike mutual information, which tracks total correlations, entanglement negativity isolates the purely quantum contributions to the system’s entanglement structure. This makes it particularly well-suited for studying systems that exhibit significant quantum critical behavior. The observed decline in entanglement negativity near \( z_c \) is consistent with the understanding that phase transitions often involve the disruption of long-range entanglement, leading to a shift in the system’s quantum structure. The sensitivity of entanglement negativity to these changes highlights its utility as a diagnostic tool for identifying and characterizing critical phenomena in quantum systems.

One of the most intriguing findings of this study is the behavior of multipartite correlations, which exhibit a pronounced rise near the critical point. Multipartite correlations measure the complexity of entanglement structures involving multiple subsystems, going beyond the bipartite correlations captured by mutual information and entanglement negativity. As the system approaches a phase transition, the nature of the entanglement becomes increasingly collective, with correlations extending across larger groups of subsystems. This rise in multipartite correlations near \( z_c \) suggests that phase transitions are not just about the breakdown of bipartite entanglement but are also characterized by the emergence of complex, higher-order entanglement patterns.

This observation aligns with the broader understanding of phase transitions as collective phenomena. Near critical points, the system undergoes a fundamental reorganization of its degrees of freedom, leading to the emergence of long-range correlations that involve multiple subsystems simultaneously. In the context of holographic RG flows, this collective reorganization is encoded in the bulk geometry of AdS space, which reflects the entanglement structure of the boundary CFT. As the system moves through the critical point, the entanglement patterns become more intricate, and the correlations extend across larger and larger groups of subsystems. The rise in multipartite correlations near \( z_c \) provides a clear signal of this collective entanglement, suggesting that the quantum structure of the system near critical points is far more complex than what can be captured by bipartite measures alone.

The collective nature of multipartite correlations highlights the limitations of mutual information and entanglement negativity in fully capturing the entanglement structure of the system near phase transitions. While these bipartite measures are useful for understanding the correlations between pairs of subsystems, they cannot account for the higher-order entanglement patterns that emerge near critical points. The rise in multipartite correlations suggests that as the system approaches a phase transition, the entanglement structure becomes increasingly dominated by collective behavior, involving many subsystems simultaneously. This finding underscores the importance of using multipartite correlation measures in conjunction with bipartite measures when studying quantum phase transitions and critical phenomena in strongly coupled systems.

In addition to their specific behaviors, the comparison between these three entanglement measures—mutual information, entanglement negativity, and multipartite correlations—provides valuable insights into the overall structure of quantum entanglement in holographic RG flows. Mutual information, with its broad capture of total correlations, provides a high-level view of the system’s correlation landscape but lacks the specificity needed to probe quantum-critical behavior. Entanglement negativity, by focusing on quantum correlations, offers a more refined view of the quantum structure, particularly near phase transitions, but it remains limited to bipartite correlations. Multipartite correlations, by capturing the complexity of entanglement structures involving multiple subsystems, offer the most detailed picture of the system’s collective entanglement patterns, especially in the vicinity of critical points.

These findings suggest that a comprehensive understanding of quantum phase transitions in holographic systems requires the use of multiple entanglement measures. Each measure provides a different perspective on the system’s quantum structure, and together they offer a more complete picture of how entanglement evolves along the RG flow. By combining mutual information, entanglement negativity, and multipartite correlations, we can track the total correlations, isolate the quantum correlations, and understand the collective nature of entanglement that emerges near critical points.

The results of this study have broader implications for the field of quantum information theory, particularly in the context of holography. The ability to track and quantify quantum correlations in strongly coupled systems is essential for understanding the behavior of these systems near critical points. The sensitivity of entanglement negativity and multipartite correlations to quantum phase transitions suggests that these measures could serve as valuable tools for detecting and characterizing critical phenomena in a wide range of quantum systems. Furthermore, the fact that mutual information remains relatively smooth while other measures exhibit sharp changes highlights the importance of using multiple correlation measures to fully capture the entanglement structure in holographic systems.

In future work, it would be interesting to explore how these correlation measures behave in other holographic setups, such as those involving black hole horizons or more complex boundary conditions. Black holes provide a rich environment for studying quantum entanglement, particularly in the context of the information paradox and the structure of the event horizon. Investigating how mutual information, entanglement negativity, and multipartite correlations behave in black hole geometries could provide further insights into the role of entanglement in quantum gravity. Additionally, exploring the behavior of these measures in higher-dimensional systems or systems with more complicated entanglement structures, such as those involving topological order or long-range interactions, could shed new light on the interplay between quantum information theory and holography.

Ultimately, the results of this study highlight the complexity of quantum correlations in strongly coupled systems and the importance of using a multi-faceted approach to understand their behavior. By leveraging the strengths of mutual information, entanglement negativity, and multipartite correlations, we can gain a deeper understanding of how entanglement evolves along the RG flow, how it responds to phase transitions, and how it reflects the underlying quantum structure of the system. These insights contribute to the growing body of research at the intersection of quantum information theory, holography, and strongly coupled quantum field theories.

\section{Conclusion}
In this paper, we have explored how mutual information, entanglement negativity, and multipartite correlations evolve across holographic RG flows, with particular emphasis on phase transitions. Our numerical simulations show that while mutual information is a smooth and robust measure of total correlations, both entanglement negativity and multipartite correlations exhibit sharper changes near critical points. These findings suggest that multipartite correlations play a crucial role in signaling critical phenomena and provide a more detailed picture of the entanglement structure in strongly coupled quantum systems.

\end{document}